%% file: main.tex
\documentclass[preprint]{article}
\usepackage{spconf}
\usepackage{amsmath, graphicx}

\usepackage[hidelinks]{hyperref}
\usepackage{csquotes}
\usepackage{cleveref}

\usepackage{booktabs}
\usepackage{tabularx}
\usepackage{makecell}

\usepackage{multirow}

\usepackage{xcolor}

\usepackage[acronym]{glossaries} 

\newacronym{dnn}{DNN}{Deep Neural Network}
\newacronym{mlp}{MLP}{Multilayer Perceptron}
\newacronym{cnn}{CNN}{Convolutional Neural Network}
\newacronym{rnn}{RNN}{Recurrent Neural Network}
\newacronym{crnn}{CRNN}{Convolutional Recurrent Neural Network}
\newacronym{asr}{ASR}{Automatic Speech Recognition}
\newacronym{ema}{EMA}{Exponential Moving Average}
\newacronym{lstm}{LSTM}{Long Short-term Memory}
\newacronym{apc}{APC}{Autoregressive Predictive Coding}

\usepackage{siunitx}						
\DeclareSIUnit\byte{B}

\title{Noise-Robust Keyword Spotting through Self-supervised Pretraining}

\name{Jacob Mørk$^{1}$, Holger Severin Bovbjerg$^{1}$, Gergely Kiss$^{1}$, Zheng-Hua Tan$^{1, 2}$} %
\address{$^{1}$Department of Electronic Systems, Aalborg University, Denmark \\
$^{2}$Pioneer Centre for AI, Denmark}
\begin{document}
\ninept
\maketitle
\begin{abstract}
Voice assistants are now widely available, and to activate them a keyword spotting (KWS) algorithm is used. Modern KWS systems are mainly trained using supervised learning methods and require a large amount of labelled data to achieve a good performance. 
Leveraging unlabelled data through self-supervised learning (SSL) has been shown to increase the accuracy in clean conditions. 
This paper explores how SSL  pretraining such as Data2Vec can be used to enhance the robustness of KWS models in noisy conditions, which is under-explored. 
 Models of three different sizes are pretrained using different pretraining approaches and then fine-tuned for KWS. 
These models are then tested and compared to models trained using two baseline supervised learning methods, 
one being standard training using clean data and the other one being multi-style training (MTR). 
The results show that pretraining and fine-tuning on clean data is superior to supervised learning on clean data across all testing conditions, and superior to supervised MTR for testing conditions of SNR above $\SI{5}{\decibel}$. 
This indicates that pretraining alone can increase the model's robustness. 
Finally, it is found that using noisy data for pretraining models, especially with the Data2Vec-denoising approach, significantly enhances the robustness of KWS models in noisy conditions.
\end{abstract}
\section{Introduction}
\label{sec:intro}
\input{incl/introduction}


The source code used to produce the results of this paper is made publicly available. 
\footnote{\href{https://github.com/aau-es-ml/ssl\_noise-robust\_kws}{https://github.com/aau-es-ml/ssl\_noise-robust\_kws}}

\section{Methodology and Data Sets}






\input{incl/Method_and_data_set}

\section{Experiments}
\input{incl/Experiments}





\section{Results}

\input{incl/Results}

\section{Conclusions}
\input{incl/conclusions}

\clearpage

\bibliographystyle{IEEEbib}
\bibliography{references}

\end{document}

%% file: incl/introduction.tex

Nowadays, voice assistants are available on almost every computer and smart device. 
Such voice assistants utilize automatic speech processing (ASR) models to transcribe human speech.
These ASR models require high computational power, which makes them infeasible to run on smaller embedded devices. 
Instead, the ASR model usually runs on a remote server and is activated by a keyword spotting (KWS) algorithm, which triggers when a specific keyword is spoken. 
KWS requires much less computation relative to ASR and can be implemented on devices with limited computation resources \cite{lopez2021deep}. 


Current state-of-the-art KWS models are based on deep neural networks \cite{lopez2021deep}, \cite{berg2021keyword}. 
These models are mainly trained in a supervised manner and require large labelled datasets to achieve good performance. 
Creating such datasets requires manual human labelling, which is time-consuming and expensive. 
Self-supervised learning methods, on the other hand, form a pseudo-target from the data itself and aim to learn a good representation of the data domain without the need for data labels \cite{jing2021understanding}, \cite{baevski_wav2vec2}. 

A recent study used the Data2Vec \cite{Baevski_data2vec} framework to pretrain a transformer-based KWS model \cite{Bovbjerg_SSL_KWS} on unlabelled data.
The study found a significant improvement in accuracy compared to a purely supervised training approach, when labelled data is limited. 
However, the study in \cite{Bovbjerg_SSL_KWS} assumes clean conditions for the audio input to the  KWS model, whereas KWS systems are usually deployed in diverse and possibly noisy environments. 

Several studies have been conducted on the noise-robustness of ASR-related tasks in self-supervised learning domains \cite{robustwav2vec}, \cite{zhu_robustdata2vec}. 
For example, \cite{zhu_robustdata2vec} utilized a Data2Vec framework combined with a contrastive loss to increase the noise robustness of a large transformer-based ASR model. 
When it comes to keyword spotting, most studies focus on supervised methods, such as multi-style training (MTR) or adversarial training \cite{lopez2021deep}, \cite{lopez2023filterbank}.
As a result, the use of self-supervised pretraining to increase the noise-robustness of keyword spotting is currently under-explored. 

In our work, we systematically investigate how self-supervised pretraining affects the robustness of the trained KWS models in noisy conditions, while we also propose some alterations to the pretraining setup that further improve the robustness of the trained KWS model. 
For example, in the Data2Vec-denoising variant, we use noisy data as input for the student branch of Data2Vec and the corresponding clean data as input for the teacher branch to simultaneously learn a good speech representation and perform denoising.
We test our models in 7 various noisy conditions at SNR levels ranging from \SIrange{-10}{20}{\decibel} in steps of \SI{5}{\decibel}, and compare different pretraining setups to purely supervised training approaches.


The results show the following:
\begin{enumerate}
    \setlength\itemsep{0.2em}
    \item Pretraining and fine-tuning on clean data yields higher accuracy than supervised training on clean data in all testing conditions.
    \item For SNR larger than \(\SI{5}{\decibel}\), clean pretraining and fine-tuning outperforms supervised training using multistyle training for both seen and unseen noise types. 
    This is interesting as the former does not use any noisy data during training. 
    \item Using noisy data for the student and clean data for the teacher in Data2Vec pretraining (i.e., Data2Vec-denoising), yields the best performing models in noisy conditions, while only performing marginally worse in clean conditions compared to models pretrained on clean data.
    \item The improvement in robustness is consistent over different model sizes. 
\end{enumerate}

%% file: incl/Method_and_data_set.tex
\subsection{Keyword spotting models}
\begin{figure}[tb]
    \centering
    \includegraphics[scale=0.55]{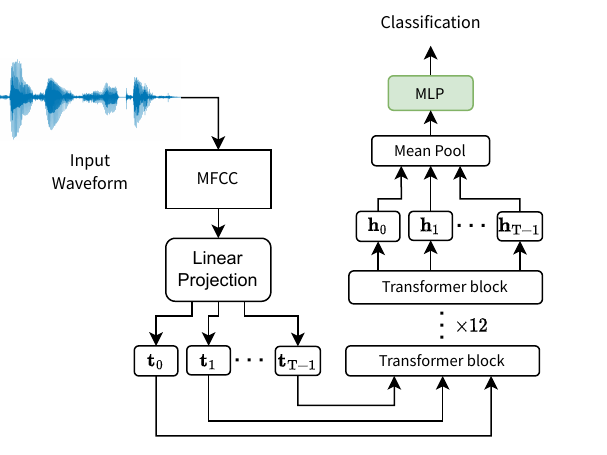}
    \caption{Illustration of the KWS system.}
    \label{fig:KWS_system}
\end{figure}

In general, deep KWS systems consist of three blocks: speech feature extraction, deep neural network (DNN) acoustic model, and posterior handling. 
The KWS system used in our work is illustrated in \Cref{fig:KWS_system}.

For feature extraction, we use Mel-frequency Cepstral Coefficients (MFCCs) \cite{davis1980comparison}. 
MFCCs are extracted using a window length of \SI{30}{\milli\second} and a hop length of \SI{10}{\milli\second}.  The extracted 40-dimensional MFCC features are then projected to the input dimension of the acoustic model through a linear layer.



Following \cite{Bovbjerg_SSL_KWS} the acoustic model is based on a Keyword Transformer (KWT) which consists of 12 Transformer blocks. 
As in the standard transformer, cosine positional encodings are added to the input before being processed in the encoder. 
The outputs of the transformer blocks for each time step are mean pooled and a multilayer perceptron (MLP) is then used for classification.

To test the influence of model size we adopt the same setup as in \cite{Bovbjerg_SSL_KWS}, varying the number of attention heads from one to three and the encoder dimension from \(64\) to \(192\), yielding three models, namely KWT-1, KWT-2, and KWT-3, with 0.6M, 2.4M, and 5.4M parameters, respectively.

\subsection{Multistyle training}
A common method to improve the robustness of a supervised model is to introduce noise when training the model. 
This can be done by adding noise to the training data, generally seen as multistyle training (MTR), and has been shown to improve the robustness of KWS models in a supervised learning setting \cite{Prabhavalkar_multistyle_training}. Therefore, we also carry out experiments applying this method during supervised training, to compare the robustness gained from MTR and self-supervised pretraining, respectively.

\subsection{Pretraining setup}
For pretraining of the KWS model, we adopt the Data2Vec \cite{Baevski_data2vec} framework, which consists of a student model and a teacher model. 
The teacher and student use identical transformer encoders, but receive different inputs.
Specifically, the teacher receives an unmasked input while the student model receives a masked version of the input. 
The goal of the student is then to predict the hidden space representation of the teacher model of the masked part of the input. 
The targets are the outputs of the top \(K\) blocks of the teacher network, corresponding to masked time steps in the student model input. 
The training target at time step \(t\), for a network with \(L\) blocks in total, is then
\begin{equation}    
    y_t = \frac{1}{K} \sum_{l=L-K+1}^L \hat{a}_t^l,
\end{equation}
where \(\hat{a}_t^l\) is the normalized output of block \(l\) at time step \(t\). 

The student model weights are updated using standard error backpropagation, with the loss being the mean squared error (MSE) between the student prediction $y'_t$ and target $y_t$. 
The weights of the teacher model are an exponential moving average (EMA) of the student model weights.
Specifically, the teacher weights are set to $\Delta:=\tau \Delta + (1 - \tau)\theta$, where $\theta$ is the student model weights, $\Delta$ is the teacher weights and $\tau$ is a smoothing factor.

\begin{figure}[ht]
    \centering
    \includegraphics[scale=0.55]{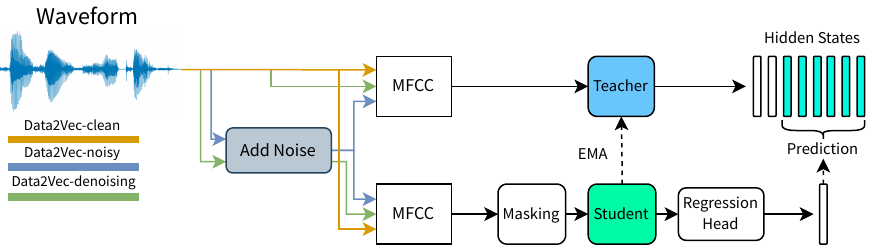}
    \caption{Illustration of the various Data2Vec pretraining setups. Here, the Data2Vec-clean input signal follows the yellow path, Data2Vec-noisy follows the blue, and Data2Vec-denoising follows the green. Black arrows denote signal paths common for all three configurations.}
    \label{fig:data2vec_setups}
\end{figure}

We also investigate two different alterations to the Data2Vec framework as depicted in \Cref{fig:data2vec_setups}, with the goal of further improving robustness to noise.
First, we simply add a data augmentation step in which the input waveform is corrupted with background noise, such that both the student and teacher receive a noisy input, denoted as Data2Vec-noisy.
Secondly, we construct a setup in which only the input for the student model is noisy, which we denote as Data2Vec-denoising. 
Since the input for the teacher model is clean, the student model will learn to denoise the input in this setup.

\subsection{Data sets}
For training and evaluation of the models, we use the Google Speech Commands V2 dataset \cite{warden2018speech}. 
The full Google Speech Commands V2 data set consists of \num{105829} relatively clean recordings of \(35\) different keywords, each with a duration of \SI{1}{\second}. The distribution of keywords in the recordings is fairly even. 
To simulate a situation with limited labelled data, we split the original training data set such that \SI{80}{\percent} of the original training data is used for unlabelled pretraining, while only \SI{20}{\percent} of the labelled data are available for supervised training.
The final split of the recordings can be seen in \Cref{tab:data_split}.

\begin{table}[ht]
    \centering
    \begin{tabular}{@{}lcccc@{}}
        \toprule
            & Pretraining & Training & Validation & Test \\ 
        \midrule
        Recordings   & 67,874  & 16,969 & 9,981 & 11,005 \\ 
        Hours & 18.9 & 4.7 & 2.8 & 3.1 \\
        \bottomrule
    \end{tabular}
    \caption{Data splits of the 105,820 recordings, and their corresponding equivalent duration in hours.}
    \label{tab:data_split}
\end{table}

In addition to the clean data set, we create a number of noise-augmented data sets. 
Here, we use six noise types, namely: bus (BUS), pedestrian (PED), street (STR), speech-shaped noise (SSN), babble (BBL), and café (CAF). BUS, PED, STR, and CAF are from the CHIME3 data set \cite{barker2015third} and BBL and SSN have been generated by \cite{kolboek2016speech}.
During training, only BUS, PED, STR and SSN are used while BBL and CAF are saved for testing as unseen noise types. 

The noisy training data set is generated by randomly adding either BUS, PED, STR or SSN noise to each individual keyword recording at an SNR level chosen uniformly from [\SI{-10}{\decibel}, \SI{-5}{\decibel}, \SI{0}{\decibel}, \SI{5}{\decibel}, \SI{10}{\decibel}, \SI{15}{\decibel}, \SI{20}{\decibel}]. 
To ensure that some clean speech appears during training, we only add noise to \SI{50}{\percent} of the training data. 
Furthermore, 42 data sets have been created for testing in various noisy conditions, one for each noise type and SNR level combination.


%% file: incl/Experiments.tex
Following \cite{Bovbjerg_SSL_KWS}, we carry out experiments for three KWT model sizes.
When using self-supervised pretraining, the models are first pretrained on the unlabelled pretraining set and then fine-tuned on the smaller labelled training set.
As a baseline for the pretrained models, we train a purely supervised model on the labelled training set.
Both baseline and pretrained models are evaluated in seen and unseen noisy conditions.
The following section describes the setup used for these experiments.

The experiments were carried out using an Nvidia A40 GPU with 48 GB RAM and 32 CPU cores available. In this setup, pretraining the largest model took 2 hours and fine-tuning them 15 minutes.


\subsection{Supervised baseline}
Using the same setup as in \cite{Bovbjerg_SSL_KWS} the supervised baseline models are trained for \(140\) epochs with a batch size of \(512\), using cross entropy as the learning objective. The weights are updated using the AdamW \cite{loshchilov2018fixing} optimizer, with a max learning rate of \(1 \cdot 10^{-3}\) and a weight decay of \(0.1\). 
The learning rate follows a linear warmup schedule for the first 10 epochs, after which a cosine annealing schedule is used. 
Furthermore, SpecAugment \cite{specaugment} is applied during training, randomly masking blocks in both time and feature dimensions. 

\subsection{Pretraining and fine-tuning}
For pretraining, the unlabelled pretraining set containing \(80\%\) of the training data is used. For masking, a Wav2Vec2 time-domain masking strategy is used \cite{baevski_wav2vec2}. 
Here, a number of MFCC vectors are randomly sampled and the following 10 MFCC vectors are replaced by a mask token embedding.
The MFCC vectors are sampled such that the overall mask probability of an MFCC vector being masked is  \(0.65\).
The other hyperparameters are identical to those used in \cite{Bovbjerg_SSL_KWS}, where most of the hyperparameters have been chosen according to the original Data2Vec study \cite{Baevski_data2vec}, with a few changes due to differences in the data sets and hardware limitations. 

After the pretraining, the models are fine-tuned using the smaller, labelled data set containing \(20\%\) of the training data. 
Fine-tuning is done using the same settings as for the supervised baseline models, however, with the transformer encoder weights initialized from a pretrained model.


\subsection{Metrics}
Classification accuracy is used as the evaluation metric, as the keyword classes are fairly evenly distributed.
The tests are carried out individually for each noise type. 
The results are averaged to calculate the average accuracy at a specific SNR. 
We compute the average accuracy for both seen and unseen noise types at each individual SNR level. 
Additionally, we evaluate the models in clean conditions.

\subsection{Training methods}
For all three model sizes, we carry out experiments on six different training methods, two supervised baselines and four Data2Vec methods. 
These are summarized in \Cref{tab:names_of_models}. 
The baselines are purely supervised methods using the \SI{20}{\percent} labelled data, with one using clean data for training and the other one using MTR training approach.

We use three different pretraining methods based on Data2Vec, namely Data2Vec-clean, Data2Vec-noisy and Data2Vec-denoising.
Data2Vec-clean conducts pretraining on clean data, whereas Data2Vec-noisy performs pretraining on noisy data for both the teacher and the student branches.
Lastly, Data2Vec-denoising uses clean data for the teacher and the corresponding noisy data for the student during pretraining, which effectively forces the student to denoise the input. 

For models pretrained with Data2Vec-clean, two finetuning approaches are applied, one conducting finetuning on clean data and the other one using MTR.
The models pretrained by Data2Vec-noisy and Data2Vec-denoising are only finetuned using MTR.

\begin{table}[ht]
    \centering
    \footnotesize
    \begin{tabular}{@{}lll@{}}
        \toprule
        Training method & Pretraining data & Finetuning data \\ \midrule
        Baseline-clean & - & clean \\ 
        Baseline-MTR & - & noisy + clean \\ 
        Data2Vec-clean & clean & clean \\
        Data2Vec-clean + noisy & clean & noisy + clean\\
        Data2Vec-noisy & noisy + clean  & noisy + clean \\
        Data2Vec-denoising  & \makecell[l]{ Teacher: clean \\ Student: noisy + clean  } & noisy + clean \\
        
        \bottomrule
    \end{tabular}
    \caption{An overview of the different  training methods. Clean refers to the data used in that particular training is unaltered. Noisy+clean refers to \SI{50}{\percent} of the data used being unaltered and the other \SI{50}{\percent} being added with a background noise, and the noise types used are: BUS, BBL, PED, and STR.}
    \label{tab:names_of_models}
\end{table}

%% file: incl/Results.tex
In this section, the results of our experiments are presented.
\Cref{tab:mean_scores} presents the mean accuracies (over noise types and SNR values including clean) of the different methods across three different KWT models,  for tests on both seen and unseen noise types.
Looking at \Cref{tab:mean_scores},
we observe that Data2Vec-denoising performs best for both seen and unseen noise, regardless of model size and the medium-sized KWT-2 model achieves slightly better performance compared to KWT-1 and KWT-3.
Our experimental results show that the average relative accuracy difference in seen noise between the best self-supervised approach, Data2Vec-denoising, and the best supervised approach, Baseline-MTR, are \(16.26\%\) for KWT-1, \(16.96\%\) for KWT-2, and \(16.35\%\) KWT-3. 
In unseen noise, we observe an improvement of \(16.22\%\) for KWT-1, \(17.45\%\) for KWT-2, and \(18.04\%\) KWT-3.
This shows a consistent and substantial improvement in robustness to noise for the pretrained models, in both seen and unseen noise and regardless of model size. 


\begin{table}[tb] 
    \centering
    \footnotesize
    \setlength\tabcolsep{1.5mm}
    \begin{tabular}{@{}lllllll@{}}
        \toprule
         & \multicolumn{2}{c}{KWT-1} & \multicolumn{2}{c}{KWT-2} & \multicolumn{2}{c}{KWT-3} \\
         & Seen & Unseen & Seen & Unseen & Seen & Unseen \\ 
         \hline
         Baseline-Clean & 0.524 & 0.532 & 0.513 & 0.521 & 0.509 & 0.502 \\
         Baseline-MTR & 0.609 & 0.592 & 0.613 & 0.596 & 0.581 & 0.560 \\
         Data2Vec-clean & 0.606 & 0.601 & 0.635 & 0.625 & 0.571 & 0.566 \\
         \makecell[l]{ Data2Vec-clean \\ + noisy } & 0.693 & 0.676 & 0.711 & 0.695 & 0.658 & 0.645 \\ 
         \makecell[l]{ Data2Vec-noisy} & 0.692 & 0.676 & 0.699 & 0.680 & 0.655 & 0.641  \\
         \makecell[l]{ Data2Vec-denoising \\ + noisy } & \textbf{0.708} & \textbf{0.688} & \textbf{0.717} & \textbf{0.700} & \textbf{0.676} & \textbf{0.661} \\ 
        \bottomrule
    \end{tabular}
    \caption{Mean accuracies of the models by averaging the accuracies at every noise level, including clean condition. The highest accuracies are highlighted with bold font.}
    \label{tab:mean_scores}
\end{table}

\begin{figure}[tb]
    \centering
    \includegraphics[width=\linewidth]{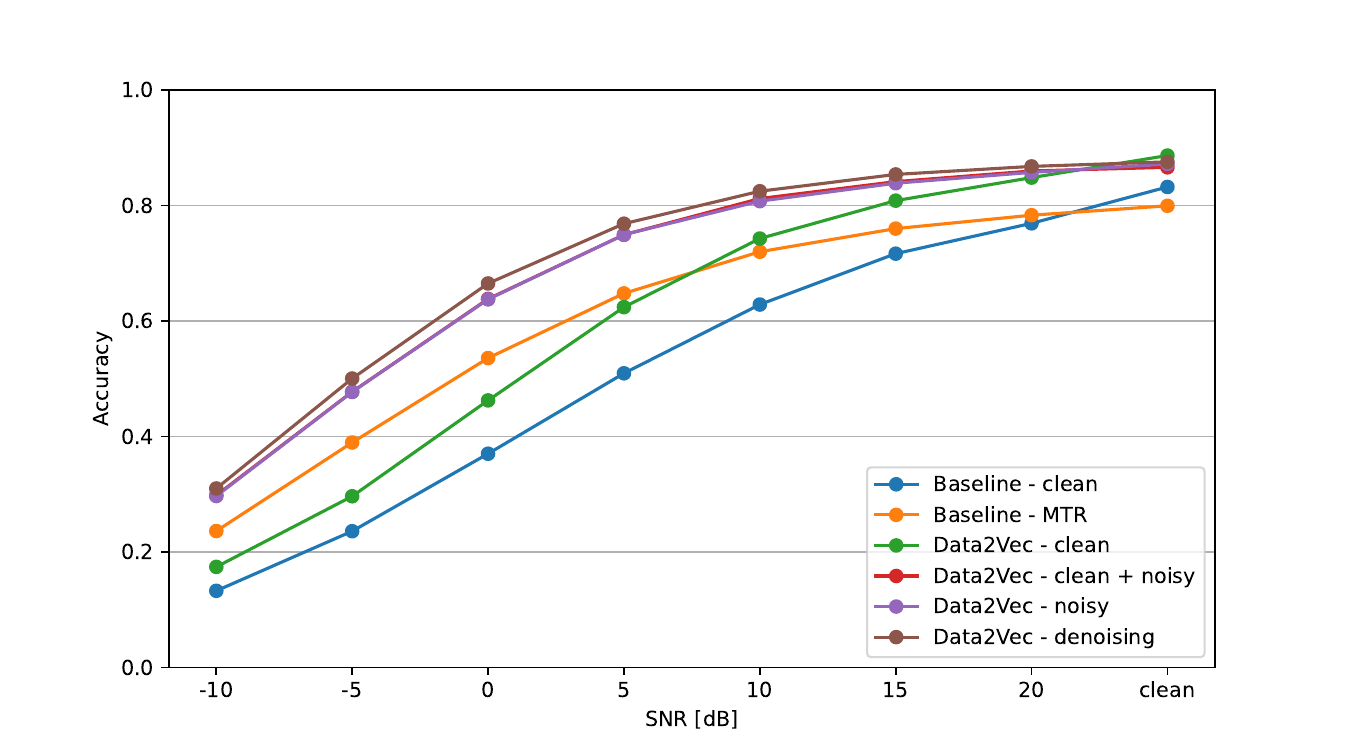}
    \caption{Visualization of the results for the KWT-1 models, tested on data with seen noise types. The results from the KWT-2 and KWT-3 models follow a similar pattern.}
    \label{fig:kwt1_known}
\end{figure}

In \Cref{tab:KWT-1_seen} the results of the KWT-1 models are presented in detail for each SNR level for seen noise. 
Additionally, they are visualized in \Cref{fig:kwt1_known}.
Here the general tendencies highlighted above are easily seen. 
From the experiments on seen noise types, we observe that Data2Vec-Clean, the model pretrained and fine-tuned only on clean data, outperforms the baseline-MTR at SNRs of \SI{10}{\decibel} and above and has the highest accuracy in clean condition. 

In seen noisy conditions, we observe that the models which have been pretrained using Data2Vec and then fine-tuned using MTR outperform the baseline models and Data2Vec-Clean model. 
We also see that Data2Vec-denoising pretraining yields the most noise robust model of the pretrained models.

\begin{table}[tb]
    \centering
    \footnotesize
    \setlength\tabcolsep{1.3mm}
    \begin{tabular}{@{}lllllllll@{}}
    \toprule
        SNR [\si{\decibel}]                                         & -10    & -5     & 0      & 5      & 10     & 15     & 20     & clean  \\ \midrule
        \makecell[l]{ Baseline- \\ Clean }          & 0.133 & 0.236 & 0.370 & 0.509 & 0.629 & 0.717 & 0.769 & 0.832 \\ 
        \makecell[l]{ Baseline- \\ MTR }            & 0.236 & 0.390 & 0.536 & 0.648 & 0.720 & 0.760 & 0.783 & 0.800 \\ 
        \makecell[l]{ Data2Vec - \\ clean }         & 0.174 & 0.297 & 0.463 & 0.624 & 0.743 & 0.808 & 0.848 & \textbf{0.887} \\ 
        \makecell[l]{ Data2Vec - \\ clean+noisy }    & 0.298 & 0.478 & 0.638 & 0.749 & 0.812 & 0.841 & 0.860 & 0.866 \\ 
        \makecell[l]{ Data2Vec - \\ noisy }    & 0.297 & 0.477 & 0.638 & 0.750 & 0.808 & 0.839 & 0.857 & 0.872 \\
        \makecell[l]{ Data2Vec - \\ denoising }     & \textbf{0.310} & \textbf{0.500} & \textbf{0.665} & \textbf{0.769} & \textbf{0.825} & \textbf{0.854} & \textbf{0.868} & 0.876 \\ \bottomrule
    \end{tabular}
    \caption{Accuracy of KWT-1 models, when tested on data with seen noises. The highest accuracies are highlighted with bold font.}
    \label{tab:KWT-1_seen}
\end{table}

\begin{figure}[tb]
    \centering
    \includegraphics[width=\linewidth]{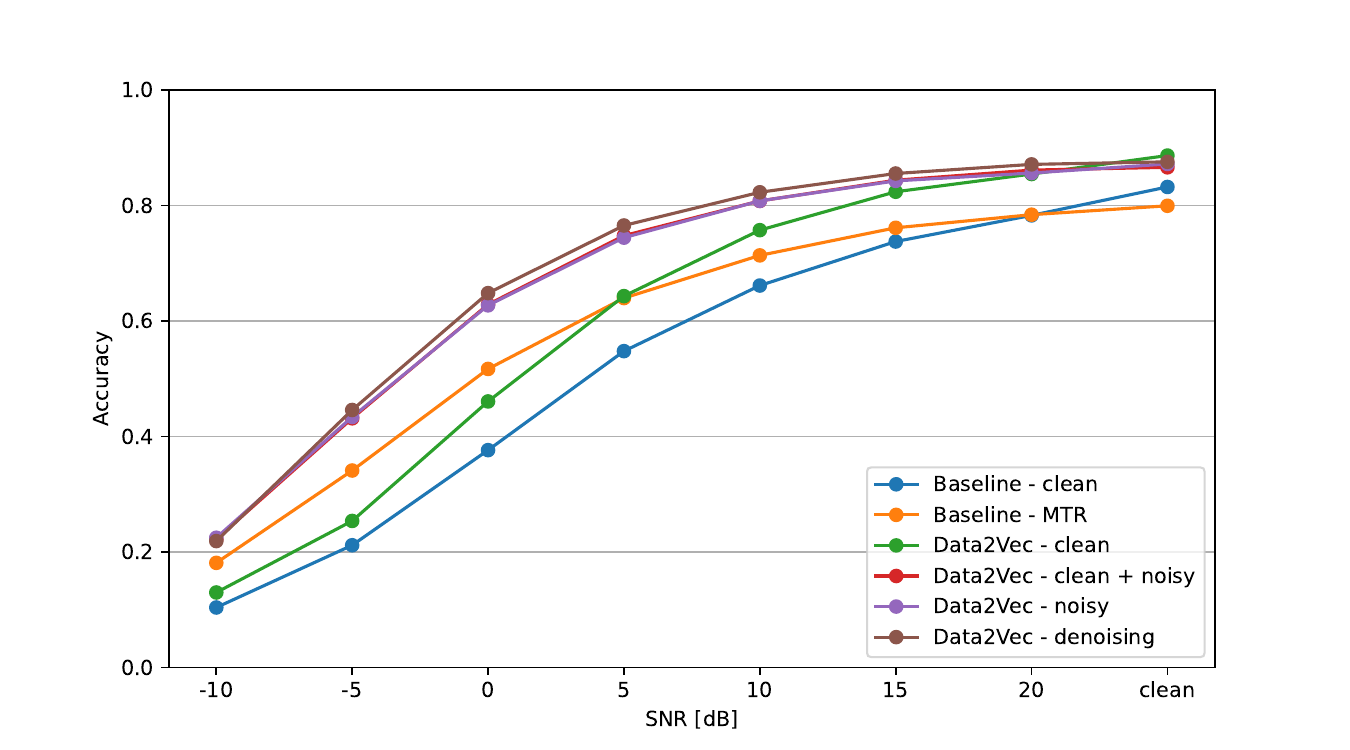}
    \caption{Visualization of the results for the KWT-1 models, tested on data with unseen noise types. The results from the KWT-2 and KWT-3 models follow a similar pattern.}
    \label{fig:all_unseen}
\end{figure}

\Cref{tab:KWT-1_unseen} shows the results of the KWT-1 models tested on unseen noise types, also visualized in \Cref{fig:all_unseen}. 
When looking at the results from testing on unseen noise types, a similar picture as for seen noise types is observed, thus the improved accuracy of the pretrained models generalizes to unseen noise types. 
Furthermore, we observe that the Data2Vec-clean model outperforms the Baseline-MTR model at SNRs above\SI{5}{\decibel}, as compared to \SI{10}{\decibel} and above for seen noise types.
This suggests that, at moderate SNR levels, the pretrained models are more robust to noisy conditions than purely supervised models, even when no noisy data is seen during pretraining or finetuning.



\begin{table}[tb]
    \centering
    \footnotesize
    \setlength\tabcolsep{1.3mm}
    \begin{tabular}{@{}lllllllll@{}}
        \toprule
        SNR                                         & -10    & -5     & 0      & 5      & 10     & 15     & 20     & clean  \\ \midrule
        \makecell[l]{ Baseline- \\ clean }          & 0.104 & 0.212 & 0.376 & 0.548 & 0.661 & 0.738 & 0.783 & 0.832 \\ 
        \makecell[l]{ Baseline- \\ MTR }            & 0.181 & 0.341 & 0.517 & 0.640 & 0.714 & 0.761 & 0.784 & 0.800 \\ 
        \makecell[l]{ Data2Vec - \\ clean }         & 0.130 & 0.254 & 0.461 & 0.643 & 0.757 & 0.824 & 0.855 & \textbf{0.887} \\ 
        \makecell[l]{ Data2Vec - \\ clean+noisy }   & 0.222 & 0.432 & 0.629 & 0.748 & 0.808 & 0.844 & 0.861 & 0.866 \\ 
        \makecell[l]{ Data2Vec - \\ noisy }         & 0.\textbf{225} & 0.434 & 0.627 & 0.744 & 0.808 & 0.843 & 0.856 & 0.872 \\
        \makecell[l]{ Data2Vec - \\ denoising }     & 0.219 & \textbf{0.446} & \textbf{0.648} & \textbf{0.765} & \textbf{0.823} & \textbf{0.855} & \textbf{0.871} & 0.876 \\ \bottomrule
 \end{tabular}
    \caption{Accuracy of the six models trained using KWT-1, when tested on data with unseen noises. The highest accuracies are highlighted with bold font.}
    \label{tab:KWT-1_unseen}
\end{table}

%% file: incl/conclusions.tex
In this paper, we investigated how self-supervised pretraining can be used as a means to make KWS models more robust against noise. 
We used the self-supervised pretraining framework Data2Vec to pretrain transformer-based KWS models of three different sizes.
After pretraining, the models are fine-tuned on a reduced version of the Google Speech Commands training set and evaluated on both clean and noisy test sets.

The results show that pretraining in general improves the robustness against noise, also when fine-tuning using MTR, for all three model sizes. 
The models which are pretrained using different methods but fine-tuned using the same MTR perform  similarly, but the Data2Vec-denoising pretraining approach yields the most robust models. 

Another observation is that the Data2Vec-clean model, which is only pretrained and trained on clean data, outperforms the Baseline-MTR model at SNRs higher than \SI{5}{\decibel} when testing on both seen and unseen noises. 
This indicates that pretraining alone increases the robustness of the model, at lower noise levels, to an extent that multistyle training cannot make up for.